\documentclass[%
 reprint,
 amsmath,amssymb,
 aps,longbibliography,
prl,
floatfix
]{revtex4-2}

\usepackage{graphicx}
\usepackage{dcolumn}
\usepackage{bm}
\usepackage{bbold}
\usepackage[dvipsnames]{xcolor}
\usepackage[colorlinks,citecolor=blue]{hyperref}

\begin{document}

\title{Motional entanglement in low-energy collisions near shape resonances}

\author{Yimeng Wang}
\author{Christiane P. Koch}%
\email{christiane.koch@fu-berlin.de}
\affiliation{%
Freie Universität Berlin, Fachbereich Physik and Dahlem Center for Complex Quantum Systems,
Arnimallee 14, 14195 Berlin, Germany
}%

\date{\today}

\begin{abstract}
Einstein, Podolsky, and Rosen discussed their paradox in terms of measuring the positions or momenta of two particles. These degrees of freedom can become entangled upon scattering, but how much entanglement can be created in this process? Here we address this question using fully coherent calculations of bipartite scattering in three-dimensional space, quantifying entanglement by the inverse of the single particle purity. We show that the standard plane-wave description of scattering fails to capture the entanglement properties, due to the essential role of quantum uncertainty in the initial state. For a more realistic description of a scattering setup, we find that the entanglement scales linearly with the scattering cross section, including strong enhancement near shape resonances, for sufficiently narrow initial momentum dispersion. We highlight the differences between scattering in one and higher spatial dimensions and discuss how the generation of motional entanglement can be detected in experiments. Our results open the way to probing, controlling, and eventually using entanglement in quantum collisions.
\end{abstract}


\maketitle

\paragraph{Introduction}
Over the past few decades, entanglement has become the cornerstone of quantum information science~\cite{NielsenChuangBook}.
The puzzling nature of entanglement as ``spooky action at a distance" is exemplarily captured by the Einstein-Podolsky-Rosen (EPR) paradox~\cite{PhysRev.47.777}. In the original EPR gedankenexperiment, it is the positions, or momenta, of two particles that are entangled. 
Such entanglement could be leveraged for encoding quantum information~\cite{hwang2025hybridqubitoscillatormodulemotional} or 
quantum-enhanced sensing~\cite{JulsgaardNature2001,PeiseNatComm2015}, but also for the 
coherent control of reactive collisions~\cite{ShapiroPRL1996,OmistePRL2018}.
The interactions required for entanglement can, for motional degrees of freedom, be leveraged through the dissociation~\cite{FryPRA1995,Opatrny2001} or photoionization~\cite{RubertiPCCPperspective2022}  of composite systems, as well as through collisional processes.
While collisions were shown to be a source of entanglement for internal degrees of freedom~\cite{BrennenPRL1999,JakschPRL1999,CalarcoPRA2000,Mandel2003,devolder2025}, their use for generating motional entanglement remains largely unexplored. 

This may in part be due to the difficulty to characterize entanglement in the infinite-dimensional Hilbert spaces of continuous variables (CV)~\cite{Eisert_2002}. 
Quantification of CV entanglement in general is limited to bi-partite pure states, using measures based on the reduced density matrices of each particle such as the von Neumann entropy~\cite{RevModPhys.81.865} or the single-particle purity~\cite{Grobe_1994}. Alternative criteria such as the positive partial transpose (PPT)~\cite{Duan2000PRL,Simon2000PRL} or quantifiers based on the Shannon entropy~\cite{Walborn2009} or the Husimi $Q$ distribution~\cite{Garttner2023} are necessary and sufficient only for Gaussian states, as encountered for photons~\cite{Serafini_Book}, but are unsuitable to quantify entanglement generated in collisions of free particles.
A second difficulty arises from the challenge of a full quantum-mechanical description of free particle collisions in three-dimensional space~\cite{Friedrich}. The generation of entanglement in collisions has been investigated only in the limit of pure $s$-wave scattering~\cite{WangPRA2006} and within one spatial dimension \cite{Opatrny2001, Law2004PRA, Tal2005PRL,Schmuser2006,Harshman2008,Benedict_2012,Schnabel2025}. 1D confinement enforces orthogonality of the transmitted and reflected amplitudes, rendering scattering in one and more spatial dimensions fundamentally distinct. Analysis of $s$-wave collisions in 3D suggested a connection between the collision cross-section and the generated entanglement, at least under conditions where the phase shift  remains nearly constant as a function of the collision energy~\cite{WangPRA2006}. However, pure $s$-wave scattering misses key features of quantum collisions --- partial wave interference and scattering resonances. The latter lead to peaks in the cross section. Will this carry over to entanglement? In other words, can scattering resonances be used to maximize the generation of entanglement in collisions? 

Here, we answer this question affirmatively. To this end, we consider elastic collisions between two distinguishable spinless particles in three-dimensional space and quantify the entanglement generated upon the collision using the inverse single-particle purity~\cite{Law2004PRA}. We show that, in the standard plane-wave description, no entanglement is generated. Solving the scattering problem for initial states with finite uncertainty, we examine the temporal evolution of entanglement and analyze scattering near a variety of shape resonances and for different geometries of the incident wave packets. Our results establish a connection between motional entanglement and the collision cross section near shape resonances, provided the initial energy dispersion is sufficiently narrow. We discuss the difference between 1D and 3D scattering, and sketch scenarios for probing the generation of entanglement in collision experiments.

\paragraph{Initial states with finite uncertainty}
To quantify the entanglement generated upon a collision, we envision a typical collision experiment using atomic or molecular beams. 
Standard quantum-mechanical scattering theory, primarily designed to calculate cross sections~\cite{Friedrich}, is only concerned with the relative motion, disregarding the center-of-mass (CM) degrees of freedom. However, when the goal is to calculate entanglement, the single-particle density matrices need to be constructed. Then the CM motion can no longer be ignored, even if relative and CM motion are not coupled. One may be tempted to assume a plane wave $e^{i\vec{P}_0\cdot\vec{R}}$ for the CM motion, as was done for initially bound particles~\cite{Tal2005PRL,Tommasini1998,Qvarfort_2020}. However, a plane wave carries no momentum uncertainty, $\Delta {P}=\sqrt{\langle P^2\rangle-\langle \vec{P}\rangle^2}=0$; and an initially separable state with $\Delta {P}=0$ (or $\Delta p=0$) will remain separable after scattering, as we show in the End Matter, cf. Eqs.\eqref{eq-371}-\eqref{eq:last}, for any non-Coulomb potential $V(r)$~\footnote{Note that $\Delta {P}=0$ has a completely different consequence for bound states, 
 as $\Delta {r}^2\cdot\Delta {P}^2=0< \Delta {r}_1^2\cdot\Delta {p}_1^2+\Delta {r}_2^2\cdot\Delta {p}_2^2$, 
implying the particles to be entangled.}.
This behavior arises from the connection between the position-momentum uncertainty and entanglement, as 
indicated by the PPT criterion~\cite{Duan2000PRL}.  
To capture the entanglement properties of 3D collisions, it is thus essential to choose a realistic initial state with finite uncertainty. A natural setting is initially spatially separated, localized particles, as prepared in collision experiments utilizing atomic or molecular beams.

\begin{figure}[tbp]
  \includegraphics[scale=0.47]{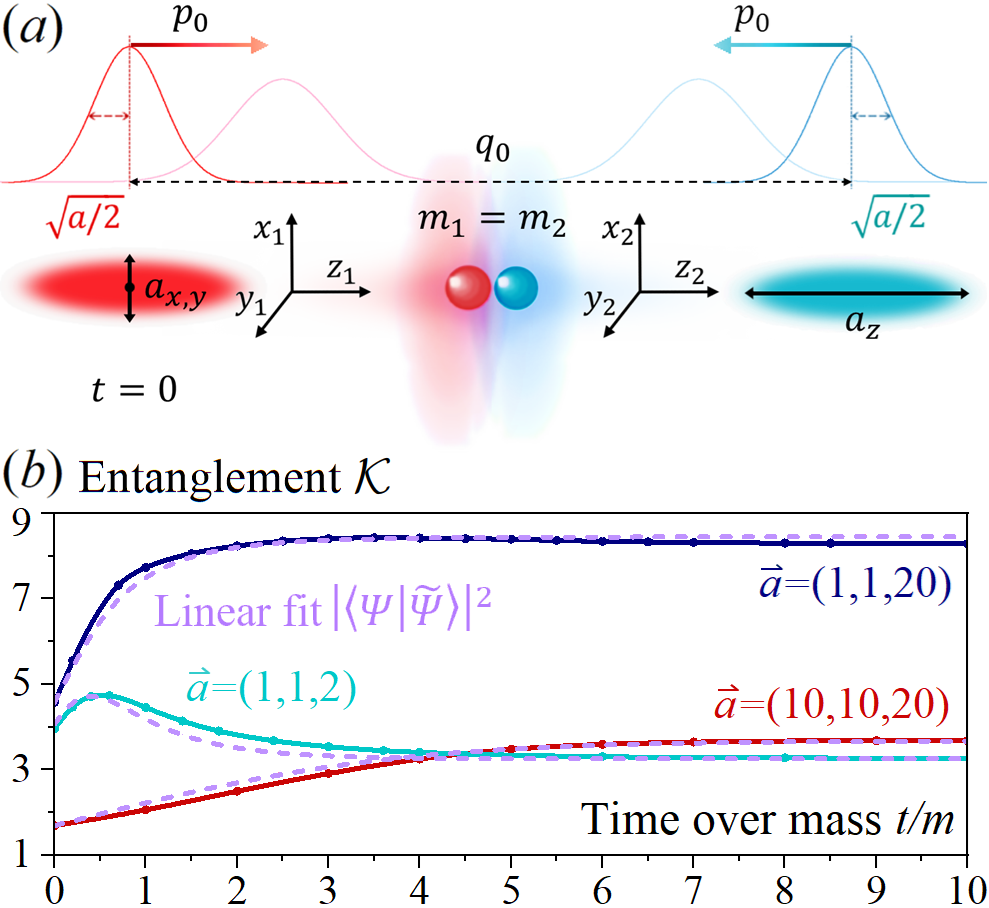}
    \caption{       (a)
    3D collision of two structureless free particles, initially with momenta $\pm{p}_{0}\hat{z}$, positions $\mp\frac{{q}_0}{2}\hat{z}$ and same spatial dispersion $\vec{a}$ that spreads under time of flight. 
    (b)
    Entanglement measure $\mathcal{K}$ for three initial Gaussian wave packets, and linear fit of $\mathcal{K}$ to the overlap of scattered and non-scattered wavefunctions versus time of flight over single particle mass 
    [$q_0=0$, $p_0=0.8$, under potential $V_0=8$ in Eq.~\eqref{eq-138}], see End Matter for the fit parameters. 
    }
  \label{f1}
\end{figure}

To model such an experiment, we consider head-on collisions of two particles with uncertainties in their kinetic energies and collimation, cf. Fig.~\ref{f1}(a). This introduces a finite dispersion in the initial relative momenta, $\Delta p^2=(1/a_x+1/a_y+1/a_z)/4$. For simplicity, we assume identical mass $m$ and opposite momenta and take the wave packets as Gaussian, so that the initial uncertainty $\Delta p\cdot\Delta r$ is minimum. The simplest potential that allows for scattering resonances is a cylindrical well with tunable width $w$ and depth $V_0$. The Hamiltonian is then given by
\begin{equation}
\label{eq-138}
    H=\frac{\vec{P}^2}{4m}+\frac{\vec{p}^2}{m}+V(r),\quad V(r)=
    \begin{cases}
        -V_0/m, & \text{} r<w   \\
        0, & \text{} r>w 
    \end{cases}\,.
\end{equation}
Unless noted otherwise, we use $w=1$.

\paragraph{Entanglement generation}
We use the inverse of the single-particle purity, $\mathcal{K}=1/\mathcal{P}$, where $\mathcal{P}=Tr(\rho_{1}^2)=Tr(\rho_{2}^2)$, to quantify the bipartite entanglement, cf. End Matter for details. When $\mathcal{K}=1$, no entanglement is generated, otherwise $\mathcal{K}>1$ (or $\mathcal{P}<1$). The finite dispersion implies a coherent superposition of different momentum components, which introduces an explicit time dependence with $\Delta r\to+\infty$ as $t\to+\infty$. The time-evolution also influences the entanglement generation as shown in Fig.~\ref{f1}(b), where we have used $q_0=0$ for clarity~\footnote{For separations $q_0>0$, it takes time for the wave-packets to start to overlap and interact, during which the behavior of $\mathcal K$ is not straightforward to interpret.}. 
Once the particles interact, $\mathcal{K}$ as a function of $t/m$ exhibits an inverse linear correlation with the overlap of the scattered ($\Tilde{\Psi}$) and non-scattered (${\Psi}$) wave packets, $F(t)=|\langle{\Psi}(t)|\Tilde{\Psi}(t)\rangle|^2=|\langle{\Psi}(0)|e^{i(H-H_0)t}\Tilde{\Psi}(0)\rangle|^2$, indicating that entanglement is generated by the interaction $V$. Wave packets with larger spatial dispersion $\vec{a}$ interact longer and therefore need more time for $\mathcal{K}$ to saturate. 
The time evolution of entanglement has largely been ignored~\cite{Tal2005PRL,WangPRA2006,Harshman2008,Schmuser2006,Weder2011, WEDER201394,Schnabel2025} which may yield incorrect values of entanglement, as discussed in the End Matter, cf. Eq.\eqref{eq-287}.

\begin{figure}[tbp]
  \includegraphics[scale=0.38]{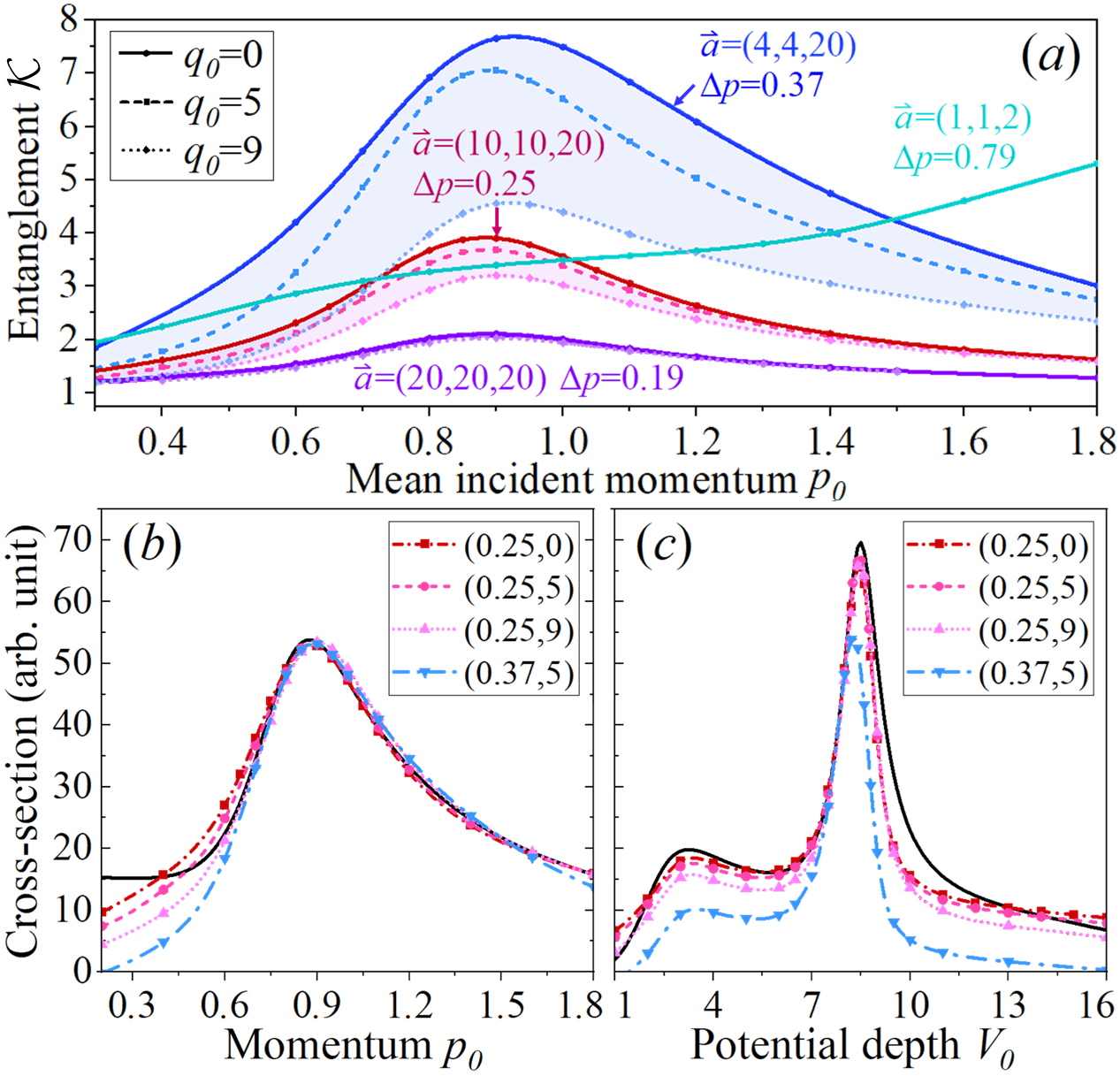}
    \caption{ 
    (a)
    Entanglement $\mathcal{K}$ vs mean incident momentum $p_0$,  near a $p$-wave shape resonance ($p_{res}=0.87$ for $V_0=8$) for different spatial (momentum) dispersions $\vec{a}$ ($\Delta {p}$) and initial locations $q_0$. 
    (b-c): 
    Cross section $\sigma$ (black solid lines) and comparison to linear fits, cf. Eq.~\eqref{eq:sigma-fit} [colored symbols with labels $(\Delta p,q_0)$] vs (b) collision momentum for $V_0=8$ and (c) potential depth with fixed $p_0=0.8$. 
    }
  \label{f2}
\end{figure}

Our subsequent analysis focuses on $\mathcal{K}(t\to+\infty)$ which characterizes the asymptotic motional entanglement, shown in Fig.~\ref{f2}(a) as a function of the mean incident momentum $p_0$ for different initial conditions. For a given spatical-momentum dispersion $\vec{a}$-$\Delta {p}$ (color-coded), a larger initial separation $q_0$ reduces $\mathcal{K}$ (cf. different line-styles) since the wave packets meet after a long free flight and therefore their transversal overlap is small in the interaction region. This reduced entanglement generation is even more pronounced for larger initial $\Delta {p}_{x,y}$. The dependence of $\mathcal K$ on the initial momentum dispersion is readily understood. In the plane wave limit, $\Delta {p}\rightarrow 0$, no entanglement is generated ($\mathcal{K}\to1$), cf. the purple curve for $\Delta {p}=0.19$ barely exceeding one. For small and intermediate $\Delta p$, the dependence of $\mathcal{K}$ on $p_0$ is reminiscent of the collision cross section, particularly near the $p$-wave shape resonance ($p_{res}=0.87$). For very broad momentum dispersion (cyan curve $\Delta {p}=0.79$, $q_0=0$), the resonance feature is completely smeared out by non-resonant momentum components. 

\paragraph{Entanglement vs cross section}
The close resemblance of the entanglement measure $\mathcal{K}$ and the plane-wave scattering cross section $\sigma$ as functions of collision energy for sufficiently small momentum dispersion $\Delta p$ in Fig.~\ref{f2}(a) suggests a linear relationship, 
\begin{equation}
\label{eq:sigma-fit}
    \sigma_{fit}(\mathcal{K})=\gamma(\mathcal{K}-\beta)\,.
\end{equation}
The agreement of $\sigma_{fit}$ with $\sigma$ is excellent, except for very low energies, cf. Fig.~\ref{f2}(b). The fit parameters $\gamma,\beta$ are reported in Table~\ref{tab:table2}. 
Our findings confirm the linear behavior found analytically in the limit of $\Delta p\to0$ for weak pure $s$-wave scattering~\cite{WangPRA2006}.
The deviation from linear behavior at low energy can be roughly explained by considering the incident momenta within the Gaussian distribution, for which $99.7\%$ of incident momenta are within $p_0\pm3\Delta p$. At low collision energy, $p_0<3\Delta p$, such that there is a discernible amount of incident momenta that propagate in the backward direction and thus do not contribute to the collision. As $p_0$ increases, more momentum components contribute to the collision, thus more entanglement is generated, explaining the monotonic increase of $\sigma_{fit}$ for small $p_0$ in Fig.~\ref{f2}(b), in contrast to the cross section which is constant. The same argument applies also to scattering with very broad $\Delta p$, for which $\mathcal{K}$ grows with $p_0$ monotonically without any resonance structure, cf. Fig~\ref{f2}(a). 

To highlight that the entanglement generated by collisions is determined by the resonance and bound-state structure, instead of the details of the potential, Fig.~\ref{f2}(c) displays the close relationship between cross section $\sigma$ and $\sigma_{fit}$ as a function of the potential depth $V_0$ (for $p_0=0.8$). When $V_0\sim 2.47$, the potential becomes deep enough to support its first $s$-wave bound state; when $V_0\sim 8$, a $p$-wave shape resonance forms, resulting in strong enhancement of both $\sigma$ and $\sigma_{fit}$. For even stronger interaction, $V_0 > 8$, both $\sigma$ and $\sigma_{fit}$ decay toward non-interacting limit, instead of increasing with $V_0$. This implies that entanglement, just as the cross-section, is determined by the scattering phase-shifts $\delta_l{(p)}$, not the short-range physics. For momentum dispersion $\Delta p=0.25$, both $\sigma$ and $\sigma_{fit}$ are maximized at $V_0=8.5$. For broader momentum dispersion $\Delta p=0.37$, $\sigma_{fit}$ maintains a similar lineshape as $\sigma$, but the maximum shifts towards $V_0=8.2$: Since a smaller $V_0$ can support shape resonances with broader widths (Table~\ref{tab:table1}), more incident momenta overlap with the resonance and therefore more entanglement is generated.

\begin{figure}[tbp]
  \includegraphics[scale=0.33]{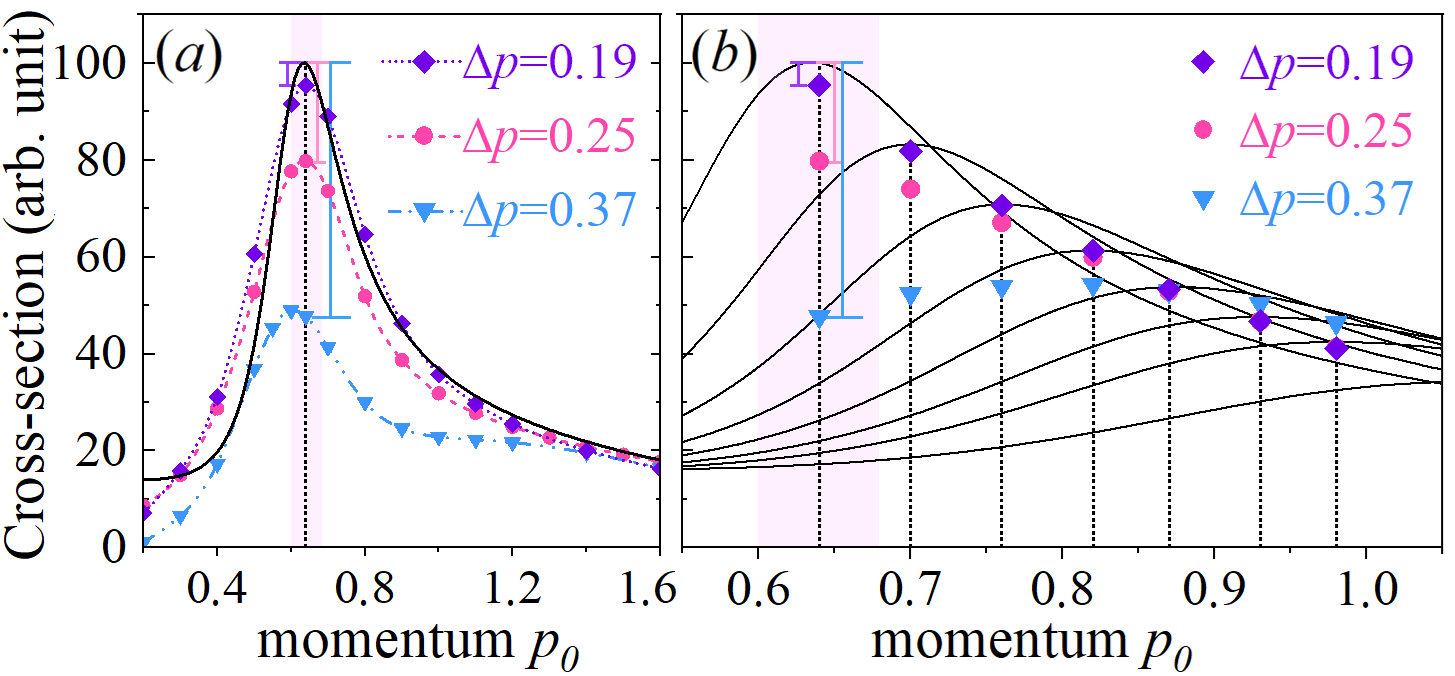} 
    \caption{ 
     Cross section $\sigma$ (black solid lines) and comparison to linear fits, cf. Eq.~\eqref{eq:sigma-fit} (colored symbols) near a $p$-wave shape resonance for (a) a single potential $V_0=8.8$ ($p_{res}=0.64$) and 
     (b)      different potentials (Table~\ref{tab:table1}) with $\sigma_{fit}$ evaluated at $p_0=p_{res}$ (vertical dashed lines),  
    for $q_0=5$, $\vec{a}=(a_{\perp},a_{\perp},20)$, $a_{\perp}=4,10,20$, cf. Fig.~\ref{f2} for the correspondence with $\Delta p$.  
    }
  \label{f3}
\end{figure}

\begin{table}[b]
\caption{\label{tab:table1}%
$p$-wave shape resonance parameters: position $p_{res}$,  resonance width $\Delta\tau$ (estimated by the FWHM of $\tau=d\delta_{(l=1)}/dE$),  cross section $\sigma_{max}$ at $p=p_{res}$ }
\begin{ruledtabular}
\begin{tabular}{cccc}
\textrm{Depth $V_0$}&
\textrm{$p_{res}$}&
\textrm{Resonance width}&
\textrm{\quad$\sigma_{max}$\quad } 
\\
\colrule %
 8.8 & 0.64 & 0.24 & 100.08 \\
 8.6 & 0.70 & 0.29 & 83.29 \\
 8.4 & 0.76 & 0.33 & 70.89 \\
 8.2 & 0.82 & 0.38 & 61.38 \\
 8.0 & 0.87 & 0.43 & 53.81 \\
 7.8 & 0.93 & 0.47 & 47.66 \\ %
 7.6 & 0.98 & 0.52 & 42.51 \\
\end{tabular}
\end{ruledtabular}
\end{table}

The role of the resonance width is studied in Fig.~\ref{f3}. Fig.~\ref{f3}(a) displays the same comparison between $\sigma$ and $\sigma_{fit}$ as Fig.~\ref{f2}(b) but for $V_0=8.8$ which supports a narrower resonance. The width of the resonance is estimated by the full width at half maximum (FWHM) $\Delta\tau$ of the Wigner time-delay $\tau=d\delta_{(l=1)}/dE$, which corresponds to the dwell time of the wave packet. Good agreement between $\sigma_{fit}$ and $\sigma$ is obtained only when the momentum dispersion $\Delta p<\Delta\tau$ [$\Delta p=0.19$ vs $\Delta\tau=0.24$ in Fig.~\ref{f3}(a)]. Otherwise, $\sigma_{fit}$ is smaller than $\sigma$, suggesting less entanglement is generated than might be expected. This is readily understood by more momentum components being non-resonant for larger $\Delta p$. For the wave packet with $\Delta p=0.25$, $\sigma_{fit}$ retains a similar shape as $\sigma$, even though the fit parameters $\gamma$, $\beta$ (which were calibrated at $V_0=8$) no longer apply here. For $\Delta p=0.37$, $\sigma_{fit}$ deviates from $\sigma$ significantly near the resonance and its peak shifts from the maximum of the cross section ($p_0=0.64$) towards the maximum of the time delay $\tau$ ($p_0\simeq0.6$). 
Since $\tau_{max}$ is a more robust indicator of resonances than $\sigma_{max}$ in the presence of interference and background scattering~\cite{Friedrich}, this deviation at larger $\Delta p$ can be attributed to the interference of the resonant and non-resonant momentum components. 
Figure~\ref{f3}(b) extends the analysis of the role of the resonance width to multiple $p$-wave shape resonances, cf. Table~\ref{tab:table1} for the resonance parameters. For $\Delta p=0.19$ which is smaller than $\Delta\tau$ for all resonances, cf. Table~\ref{tab:table1}, $\sigma_{fit}$ and thus $\mathcal{K}(p_{res})$ show the same dependence as $\sigma_{max}$ versus $p_0$. For broader $\Delta p$ (pink circles and blue triangles in Fig.~\ref{f3}(b)), $\sigma_{fit}$ drops below $\sigma$, as the narrowness of resonance reduces entanglement generation. For large $\Delta p$ ($\Delta p=0.37$), the largest value of $\sigma_{fit}$ (and thus $\mathcal K$) does not occur for the strongest resonance (at $p_{res}=0.64$) which is also the narrowest, but rather at $p_{res}=0.82$, a resonance whose FWHM $\Delta\tau\simeq\Delta p$. We thus find that entanglement generation is dominated by the shape resonance, provided the initial momentum dispersion is not too broad; otherwise the presence of too many non-resonant momentum components impedes resonance enhancement of the entanglement generation.

\begin{figure}[tbp]
  \includegraphics[width=0.95\linewidth]{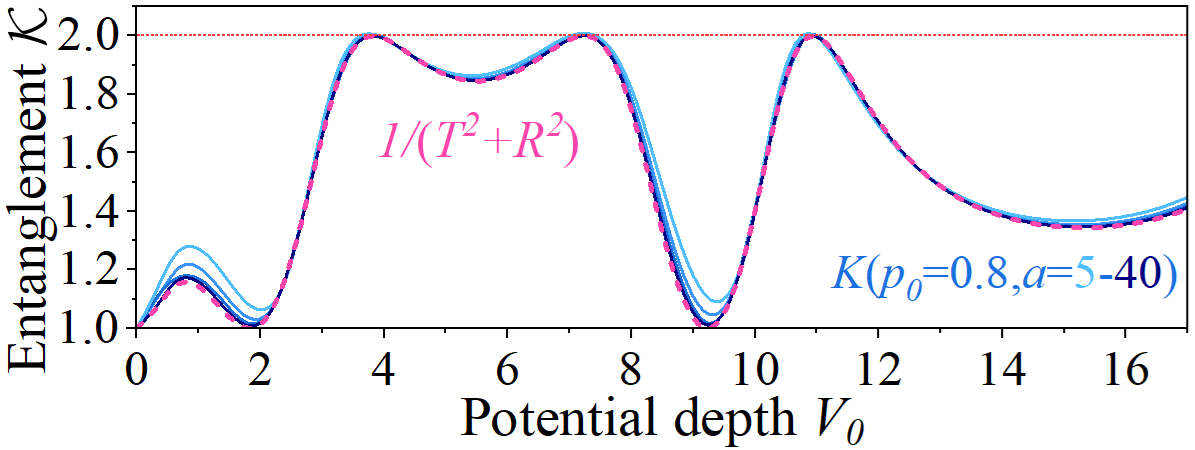}
    \caption{ 
    Entanglement in 1D scattering: $\mathcal{K}(t\to+\infty)$ vs potential depth $V_0$, calculated with $p_0=0.8$ and $a=5,10,20,40$ (solid curves with color darkening as $a$ increases). $\mathcal K$ converges onto $1/(T^2+R^2)$ (dashed pink curve, $T+R=1$). 
    }
  \label{f4}
\end{figure}

\paragraph{Comparison to 1D Scattering}
To emphasize the difference between scattering in 1D and 3D, Fig.~\ref{f4} shows the entanglement when the motion is restricted to the $z$-direction. 
In this case, the initial separation $q_0$ does not influence the asymptotic entanglement, as there is no more transversal uncertainty. 
Comparing Fig.~\ref{f4} with Fig.~\ref{f2}(c), the differences are striking. 
For $\Delta p\to0$, no entanglement is generated in 3D collisions. In contrast, in 1D, $\mathcal K$ is predicted by the transmission ($T$) and reflection ($R$) coefficients evaluated at $p_0$ (darkblue line in Fig.~\ref{f4}), reaching $\mathcal{K}_{max}=2$ when $R=T=1/2$.
For finite dispersion $\Delta p<p_0/3$ (lightblue lines), $\mathcal{K}$ still resembles $1/(T^2+R^2)$. 
Entanglement generated in a 1D collision is reminiscent of a beam splitter 
in quantum optics~\cite{Asboth2005}, as it creates superpositions of two orthogonal states, the left- and right-propagating wave-packets $|L\rangle$ and $|R\rangle$, and therefore generates maximum etanglement of $2$, analogously two qubits. The same relation $\mathcal{K}\approx1/(T^2+R^2)$ is also obtained for unequal masses~\cite{Benedict_2012}. 
In 3D in contrast, we find $\mathcal K$ larger than 2, cf. Fig.~\ref{f1}(b), which can be attributed to superpositions involving many (in principle infinitely many) single-particle $\ell_1$, $\ell_2$ partial waves. 

\paragraph{Prospects for probing entanglement generation in collision experiments}
Observables that reveal entanglement through measurable quantities are referred to as entanglement witnesses~\cite{RevModPhys.81.865}. Unfortunately, the witnesses commonly used for continuous variables in quantum information cannot successfully identify entanglement in scattering states. For example, the PPT criterion requires a decrease in uncertainty when evolving from separable to entangled states~\cite{Duan2000PRL,Simon2000PRL}, but both $\Delta r$ and $\Delta p$ increase upon scattering. The most direct approach would be to probe the single-particle reduced density matrix~\cite{RevModPhys.81.299} to extract $\mathcal{K}$. Velocity map imaging (VMI) provides access to the momentum space wavefunction~\cite{Eppink1997}; and combining it with ionization allows for probing elastic collisions~\cite{PaliwalNatComm21,PaliwalNatureChem21}. However, VMI yields the amplitudes but not the phases of the momentum space wavefunction. The latter are needed for full quantum state tomography and require an interferometric approach. This can be realized with pump-probe spectroscopy, as recently demonstrated for electronic motion~\cite{Morrigan2023PRL, Laurell2025}. In order to detect motional entanglement, pump and probe excitation need to couple to free motion of the particle. While such coupling exists (it is at the basis of laser cooling),  the momentum changes are likely too small to be measurable in a combination of an interferometric setup with VMI. Instead of reconstructing the full single-particle density matrices, a more practical approach may be to infer the entanglement generated upon a collision by comparing wave packet widths measured in coincidence and single-particle detection, as suggested for electron-ion entanglement in photoionization~\cite{Fedorov2004} and atom-photon entanglement in spontaneous emission~\cite{Fedorov2005}. To this end, the original proposal~\cite{Fedorov2004,Fedorov2005} needs to be adapted to smaller mass differences. For the experimental implementation, the measurement of elastic collisions~\cite{PaliwalNatComm21, PaliwalNatureChem21} would need to be combined by  coincidence detection with double-VMI~\cite{PitzerScience13,Margulis2020,MargulisSci23}. While clearly challenging, it requires combination of existing experimental technology. 

\paragraph{Conclusions}
We have analyzed motional entanglement generated upon the collision of two spinless particles. Collisions in one- and three-dimensional space are strikingly different, and this reflects also in the generation of entanglement.
While standard scattering theory is based on plane waves, we have found a time-dependent treatment starting from separated wave packets with finite position and momentum uncertainties to be crucial: It avoids violation of energy conservation and captures the generation of entanglement, including the role of interferences correctly. Quantifying entanglement via the inverse single-particle purity after scattering, we find a linear relationship between the asymptotic entanglement and the collision cross section for sufficiently narrow initial momentum dispersion.
We predict that scattering resonances maximize entanglement generation, unless the momentum dispersion is much broader than the width of the resonance. 
As the momentum dispersion increases, becoming comparable to the resonance width, the collision energy that maximizes entanglement shifts from the peak of the cross section to the peak of the Wigner time delay. 
Future work should extend this framework to particles with internal structure, enabling investigations into entanglement swapping across diverse degrees of freedom and the generation of multi-partite entanglement in collisions. Furthermore, generalizing this approach to inelastic and reactive collisions will provide a quantitative assessment of the role of entanglement in chemical reaction dynamics, a widely conjectured connection~\cite{Devolder2020,Devolder2021,devolder2025} that has yet to be quantified. Finally, our work highlights the need for establishing concrete protocols for experimental measurement of the entanglement in collisions, as this  will be  essential for validating any theoretical prediction.

\begin{acknowledgments}
\paragraph{Acknowledgments}
We would like to thank Ed Narevicius for helpful discussions.
Financial support by the Alexander von Humboldt Foundation is gratefully acknowledged.
We also would like to thank the HPC service and the Department of Physics at Freie Universität Berlin for computing time at the Curta and Sheldon clusters. %
\end{acknowledgments}

\bibliographystyle{apsrev4-2}

\bibliography{collisions}


\section*{End Matter}
\appendix

\paragraph{Calculation of the single-particle purity} \label{sec:a1}
The single-particle purity $\mathcal{P}=1/\mathcal{K}=Tr(\rho_{1}^2)$ is obtained as
\begin{widetext}
\begin{eqnarray}
\label{eq-401}
    \mathcal{P}(t)&=& \int d^3 {r}_1 d^3 {r}_2 d^3 {r}_1^{\,\prime} d^3 {r}_2^{\,\prime}\,\,\Tilde{\Psi}^*(\Vec{r}_1,\Vec{r}_2^{\,\prime},t)\Tilde{\Psi}^*(\Vec{r}_1^{\,\prime},\Vec{r}_2,t)\Tilde{\Psi}(\Vec{r}_1,\Vec{r}_2,t)\Tilde{\Psi}(\Vec{r}_1^{\,\prime},\Vec{r}_2^{\,\prime},t)  \,,
\end{eqnarray}     
where 
$\rho_1(\Vec{r}_1,\Vec{r}_1^{\,\prime},t)=
\int d^3r^{\,\prime}_2\,\Tilde{\Psi}^*(\Vec{r}_1,\Vec{r}_2^{\,\prime},t)\Tilde{\Psi}(\Vec{r}_1^{\,\prime},\Vec{r}_2^{\,\prime},t)$  and $\Tilde{\Psi}(\Vec{r}_1,\Vec{r}_2,t)$ is the scattered wave packet.
It differs from the unscattered wave packet $\Psi(\Vec{r}_1,\Vec{r}_2,t)$
only in the relative motion $\Vec{r}=\Vec{r}_1-\Vec{r}_2$.
Expanding $\Psi(\Vec{r}_1,\Vec{r}_2,t)$ 
into the plane-wave basis with expansion coefficients $\psi_{0\alpha}(\Vec{p}_{\alpha},t)$ 
and rotating from separate particle coordinates into CM and relative coordinates (using fractional masses $\mu_{\alpha}$ with $\mu_1+\mu_2=1$), 
\begin{eqnarray*}
    \Psi(\Vec{r}_1,\Vec{r}_2,t)&=&\int d^3 {P}\int d^3 {p}\,\,\psi_{01}(\mu_1\Vec{P}+\Vec{p},t)\psi_{02}(\mu_2\Vec{P}-\Vec{p},t) e^{i\Vec{P}\cdot(\mu_1\Vec{r}_1+\mu_2\Vec{r}_2)} e^{i\Vec{p}\cdot(\Vec{r}_1-\Vec{r}_2)}  \\
    \Tilde{\Psi}(\Vec{r}_1,\Vec{r}_2,t)&=&\int d^3 {P}\int d^3 {p}\,\,\psi_{01}(\mu_1\Vec{P}+\Vec{p},t)\psi_{02}(\mu_2\Vec{P}-\Vec{p},t) e^{i\Vec{P}\cdot(\mu_1\Vec{r}_1+\mu_2\Vec{r}_2)} f(\Vec{p},\Vec{r}_1-\Vec{r}_2) 
\end{eqnarray*}     
\end{widetext}
where $f(\Vec{p},\Vec{r})$ are the eigenfunctions of the full scattering Hamiltonian $H_0+V(r)$. Assuming that scattering is an adiabatic process,  the expansion coefficients remain unchanged upon scattering.
For any isotropic potential $V(r)$ decaying no slower than $1/r^{2}$, $f(\Vec{p},\Vec{r})$ can be expanded into partial wave components,  
\begin{equation}
\label{eq-272}
   f(\Vec{p},\Vec{r})=\sum_{l=0}^{\infty} (2l+1) i^l \psi_l(p,r) P_l(\hat{p}\cdot\hat{r}), 
\end{equation}
where $P_l$ is a Legendre polynomial and
\[
\psi_l(p,r)\xrightarrow{r\rightarrow\infty} \frac{1}{2pr} [e^{2i\delta_l(p)}e^{i(pr-\frac{l+1}{2}\pi)}+e^{-i(pr-\frac{l+1}{2}\pi)}] \,.
\]
For large $l$, $\delta_l(p)\rightarrow 0$ due to the high centrifugal barrier; for the finite range potentials in Table~\ref{tab:table1}, the cutoff $L=6$ is sufficient in the calculations. 

We take the single-particle momentum distributions $\psi_{0\alpha}(\Vec{p}_{\alpha},t)$ ($\alpha=1,2$) to be Gaussians (with $\hbar=1$), 
\begin{widetext}
\begin{equation}
\label{eq-322}
    \psi_{0\alpha}(\Vec{p}_{\alpha},t)=\left[\frac{\det A}{(4\pi^3)^3}\right]^{\frac{1}{4}} \exp\left[{-\frac{1}{2}(\Vec{p}_{\alpha}-\Vec{p}_{0\alpha})^{T} A(\Vec{p}_{\alpha}-\Vec{p}_{0\alpha})}\right] \exp\left[{-\frac{it}{2m}p_{\alpha}^2}\right] \exp\left[{i\Vec{p}_{\alpha}\cdot\Vec{q}_{0\alpha}}\right]  \,,
\end{equation} 
\end{widetext}
assuming identical mass $m$ and dispersion $A=\mathrm{diag}(a_x,a_y,a_z)$, 
such that the scattered state $\Tilde{\Psi}$ is separable in the CM and relative coordinates. 
We integrate Eq.~\eqref{eq-401} using Gauss-Legendre quadratures along each direction of $\vec{r}_{\alpha}$. $\mathcal{K}$ is stabilized over the choice of mesh with at least two significant digits, with normalization accuracy $\langle\Tilde{\Psi}|\Tilde{\Psi}\rangle-1<10^{-2}$.  
The post-scattering entanglement $\mathcal{K}(t/m)$ shows an inverse linear correlation with the overlap $F=|\langle\Psi|\Tilde{\Psi}\rangle|^2$ as shown in Fig.~\ref{f1}(b), fitted as $-6.58\,F+6.57$, $-12.07\,F+8.59$, and $-6.72\,F+4.77$ for $\vec a=(10,10,20)$, $(1,1,20)$ and $(1,1,2)$, respectively.
The saturated entanglement $\mathcal{K}(t\to+\infty)$ is calculated at $t/m\approx20+q_0/(2\,p_0)$, and the fit parameters for the cross-sections $\sigma_{fit}$ in Figs.~\ref{f2} and~\ref{f3} are given in Table~\ref{tab:table2}. 
\begin{table}[bt]
\caption{\label{tab:table2}%
Parameters of the fit $\sigma_{fit}(\mathcal{K})=\gamma(\mathcal{K}-\beta)$ for the initial conditions in Figs.~\ref{f2}(b,c) and~\ref{f3}. 
}
\begin{ruledtabular}
\begin{tabular}{cccccc}
\textrm{$\vec{a}$}&
\textrm{(20,20,20)}&
\textrm{(10,10,20)}&
\textrm{(10,10,20)}&
\textrm{(10,10,20)}&
\textrm{(4,4,20)}   
\\
\textrm{$q_0$}&
\textrm{$q_0=5$}&
\textrm{$q_0=9$}&
\textrm{$q_0=5$}&
\textrm{$q_0=0$}&
\textrm{$q_0=5$}   
\\
\colrule %
 $\gamma$ & 49.3 & 23.3 & 18.1 & 16.2 & 9.16\\
 $\beta$ & 1.00 & 0.91 & 0.74 & 0.65 & 1.25\\
\end{tabular}
\end{ruledtabular}
\end{table}

\paragraph{Momentum-Space Analysis} 
Since the wave packet is expanding and oscillating in real space but not in momentum space, on first glance it might seem easier to calculate the entanglement in momentum space~\cite{Tal2005PRL,WangPRA2006,Harshman2008,Weder2011, WEDER201394}, 
\begin{widetext}
\begin{equation}
\label{eq-329}
\Tilde{\Psi}(\Vec{p}_1,\Vec{p}_2,t)=\int d^3 {p}\,\,\psi_{01}\left[\mu_1(\Vec{p}_1+\Vec{p}_2)+\Vec{p},t\right]\psi_{02}\left[\mu_2(\Vec{p}_1+\Vec{p}_2)-\Vec{p},t\right]
    \,\mathcal{F}(\Vec{p},\mu_2\Vec{p}_1-\mu_1\Vec{p}_2) \,,
\end{equation} 
where $\psi_{0\alpha}(\Vec{p}_{\alpha},t)$ ($\alpha=1,2$) are the Gaussians from Eq.~\eqref{eq-322}, 
$\mathcal{F}(\Vec{p},\Vec{p}_{12})$ is the Fourier transform of $f(\Vec{p},\Vec{r})$, cf. 
Eq.~\eqref{eq-272}, and $\vec p_{12}=\mu_2\Vec{p}_1-\mu_1\Vec{p}_2$ is the relative momentum after scattering. Importantly,
$\mathcal{F}(\Vec{p},\Vec{p}_{12})=\delta(\Vec{p}-\Vec{p}_{12})$ only when no scattering occurs, 
otherwise
\begin{equation}
\label{eq-287}
    \mathcal{F}(\Vec{p},\Vec{p}_{12}) =\sum_{l=0}^{\infty} \frac{(2l+1)}{8\pi^2}P_l(\hat{p}\cdot\hat{p}_{12}) \left\{
    \frac{\pi\delta(p-p_{12})}{p_{12}^2} (e^{2i\delta_l(p)}+1)+\frac{i}{p_{12}p}\textsl{P}\left[\frac{1}{p-p_{12}}+\frac{(-1)^{l+1}}{p+p_{12}}\right] (e^{2i\delta_l(p)}-1)\,.
    \right\}    
\end{equation}
\end{widetext}
The principal value term $\textit{P}\,[...]$ appears in both 1D and 3D scattering~\cite{Dirac1927}, but is ignored when treating scattering as a ``reflection" of relative momentum~\cite{Schmuser2006,Harshman2008}, which effectively assumes $|\Vec{p}|=|\Vec{p}_{12}|$. This assumption results in a time-dependency of $\exp\,{[it{(p_1^2+p_2^2)}/{2m}]}$ in $\Tilde{\Psi}(\Vec{p}_1,\Vec{p}_2,t)$, which yields a time-\textit{independent} $\mathcal{K}_{s}$. 
However, according to our calculations, $\mathcal{K}_s$ can be very different from the correct asymptotic value $\mathcal{K}(t\to+\infty)$~\footnote{Neglecting the time-dependence may also explain claims of $\mathcal{K}_{max}$ larger than 2 in 1D collisions, as entanglement overall decays with time in 1D, converging to the bound expected for two orthogonal states.}. 
Moreover, neglecting $\textit{P}\,[...]$ results in a violation of energy conservation $\langle\Tilde{\Psi}|H|\Tilde{\Psi}\rangle$. It also eliminates the relative phase-shifts between the incoming and outgoing partial wave components, i.e., a key feature of quantum scattering.

It is challenging to properly integrate over $\Vec{p}$ in Eq.~\eqref{eq-329}: The singularity near $p_{12}$ in $\textit{P}\,[...]$ is usually handled by the residue theorem, assuming the remaining integrand is analytic and vanishes at large distance. However, exponential factors as those in the Gaussians $\psi_{0\alpha}(\Vec{p}_{\alpha},t)$ diverge over half of the complex plane, impeding the common construction of a closed contour $C$. Moreover, the analyticity of the phase-shifts $\delta_l(p)$ over the complex plane can be problematic. All of this makes evaluating $\Tilde{\Psi}(\Vec{p}_1,\Vec{p}_2,t)$ rather tricky, and entanglement calculations based on momentum space need to be carefully scrutinized. 
To avoid these issues and fully capture coherence, we have evaluated entanglement in real space through $\Tilde{\Psi}(\Vec{r}_1,\Vec{r}_2,t)$, as described above. Nevertheless, Eq. \eqref{eq-287} remains instructive for elucidating why plane-wave initial states fail to generate entanglement, as we explain next. 

\paragraph{Why plane waves preclude the generation of entanglement}
Assume that the initial state satisfies any two of the following three conditions: (1) $\Delta {P}=0$. (2) $\Delta {p}=0$. (3) The particles are separable prior to collision. Then the pre-scattering state takes the form 
\begin{widetext}
\begin{equation}
\label{eq-371}
    \psi_{01}(\Vec{p}_1,t)\psi_{02}(\Vec{p}_2,t)=\frac{1}{{N}}e^{-iE_0t}\delta(\Vec{p}_1-\Vec{p}_{01})\delta(\Vec{p}_2-\Vec{p}_{02})=\frac{1}{{N}}e^{-iE_0t} \delta(\Vec{p}_1+\Vec{p}_2-\Vec{P}_{0})\delta(\mu_2\Vec{p}_1-\mu_1\Vec{p}_2-\Vec{p}_{0}) \,,
\end{equation}
where  $N$ is a normalization factor, $\Vec{p}_{0\alpha}$ is the initial momentum for particle $\alpha$, and $\Vec{p}_{0}=\mu_2\Vec{p}_{01}-\mu_1\Vec{p}_{02}$ and $\Vec{P}_{0}=\Vec{p}_{01}+\Vec{p}_{02}$ are the relative and CM initial momenta, $E_0={p_{01}^2}/{(2m_1)}+{p_{02}^2}/{(2m_2)}$ is the scattering energy (here, we make no assumptions on the particle masses $m_{1,2}$). 
The scattering wave-packet $\Tilde{\Psi}$ and the single-particle purity then become
\begin{eqnarray}
&\Tilde{\Psi}&(\Vec{p}_1,\Vec{p}_2,t)
=\frac{1}{N}e^{-iE_0t} \delta(\Vec{p}_1+\Vec{p}_2-\Vec{P}_{0})\mathcal{F}(\Vec{p}_{0},\mu_2\Vec{p}_1-\mu_1\Vec{p}_2) \,, \\
&\mathcal{P}&=(2\pi)^{12}\int d^3 {p}_1 d^3 {p}_2 d^3 {p}_1^{\,\prime} d^3 {p}_2^{\,\prime}\,\,\Tilde{\Psi}^*(\Vec{p}_1,\Vec{p}_2^{\,\prime},t)\Tilde{\Psi}^*(\Vec{p}_1^{\,\prime},\Vec{p}_2,t)\Tilde{\Psi}(\Vec{p}_1,\Vec{p}_2,t)\Tilde{\Psi}(\Vec{p}_1^{\,\prime},\Vec{p}_2^{\,\prime},t)=\frac{(2\pi)^{9}}{N^3} \int d^3 \Vec{p}\, |\mathcal{F}(\Vec{p}_{0},\Vec{p})|^4  \,,
\end{eqnarray}     
where $\mathcal{F}(\Vec{p}_{0},\Vec{p})$, cf. Eq.~\eqref{eq-287}, can be decomposed into non-scattered and a scattered components, $\mathcal{F}(\Vec{p}_{0},\Vec{p}) =\delta(\Vec{p}_{0}-\Vec{p})+\Tilde{\mathcal{F}}(\Vec{p}_{0},\Vec{p})$ with 
\begin{equation}  
    \delta(\Vec{p}_0-\Vec{p})=\sum_{l=0}^{\infty} \frac{(2l+1)}{4\pi}P_l(\hat{p}_0\cdot\hat{p}) \frac{\delta(p_0-p)}{p^2}, \quad
    \Tilde{\mathcal{F}}(\Vec{p}_0,\Vec{p})=\sum_{l=0}^{\infty} \frac{(2l+1)}{8\pi}P_l(\hat{p}_0\cdot\hat{p}) (e^{2i\delta_l(p_0)}-1)\left\{    \frac{\delta(p_0-p)}{p^2} +\textit{P}\left[...\right]     \right\}.    
\end{equation}
\end{widetext}
For $\Vec{p}_{0}=\Vec{p}$, the first term dominates, 
$\mathcal{F}(\Vec{p}_{0},\Vec{p}) =\delta(\Vec{p}_{0}-\Vec{p})\gg \Tilde{\mathcal{F}}(\Vec{p}_{0},\Vec{p})$, as $\delta_{l\to\infty}(p_0)\to 0$. Otherwise $\mathcal{F}(\Vec{p}_{0},\Vec{p}) =\Tilde{\mathcal{F}}(\Vec{p}_{0},\Vec{p})$. Therefore, separating the integral $\int d^3 \Vec{p}$ into contributions for $\Vec{p}=\Vec{p}_{0}$ and $\Vec{p}\neq\Vec{p}_{0}$ yields
\begin{equation}\label{eq:last}
\begin{split}
    \mathcal{P}&=
    1+\frac{(2\pi)^{9}}{N^3} \int_{\Vec{p}\neq\Vec{p}_{0}} d^3 \Vec{p} \,|\Tilde{\mathcal{F}}(\Vec{p}_{0},\Vec{p})|^4 
    \geq 1\,.
\end{split}
\end{equation}
Since $\mathcal{P}\leq 1$ by definition, 
Eq.~\eqref{eq:last} implies $\mathcal{P}=1$,  
showing that no entanglement is generated ($\mathcal{K}=1$). 

\end{document}